\documentclass[journal]{IEEEtran}
\usepackage[ruled,vlined]{algorithm2e}
\usepackage{cite}
\usepackage{graphicx}
\usepackage{array}
\usepackage{color}
\usepackage{amsfonts,amssymb}
\usepackage{enumitem}
\usepackage{multicol}
\usepackage{multirow}
\usepackage{nomencl}
\usepackage{import}
\graphicspath{{images/}}
\usepackage{amsmath,amsthm}
\usepackage{algorithmic}
\usepackage{url}
\usepackage{tikz} 
\usetikzlibrary{arrows,calc}
\newtheorem*{remark}{Remark}
\usepackage{subcaption}
\usepackage{empheq}
\usepackage{comment}

\DeclareMathOperator{\Tr}{Tr}
\DeclareMathOperator{\st}{s.t.}
\usetikzlibrary{arrows.meta}
\usetikzlibrary{positioning,fit,backgrounds}

\usepackage{xcolor}
 \usepackage{ifdraft}
 
  \tikzset{%
  >={Latex[width=2mm,length=2mm]},
            base/.style = {rectangle, rounded corners, draw=black,
                           minimum width=3cm, minimum height=1cm,
                           text centered, font=\sffamily},
  activityStarts/.style = {base, fill=black!30},
       startstop/.style = {base, fill=green!25},
    activityRuns/.style = {base, fill=pink!25},
         process/.style = {base, minimum width=4cm, fill=orange!25,
                           font=\sffamily},
            a/.style = {rectangle, fill = yellow!10
                           minimum width=2cm, minimum height=1cm,
                           text centered, font=\sffamily},
}

\begin{document}

\title{Degradation-Infused Energy Portfolio Allocation Framework: Risk-Averse Fair Storage Participation
\vspace{-2mm}}

\author{\IEEEauthorblockN{Parikshit Pareek$^{1 \#}$, L. P. Mohasha Isuru Sampath$^{2 \#}$, Anshuman Singh$^{3 \#}$, Lalit Goel$^3$,
        Hoay Beng Gooi$^3$,
        \\ and Hung Dinh Nguyen$^{3 \star}$ \vspace{-4mm}}
\thanks{$^\#$ Authors contributed equally in this work: P. Pareek \& L. P. M. I. Sampath}
\thanks{$^1$ is with Theoretical Division (T-5), Los Alamos National Lab, NM, USA. Email: pareek@lanl.gov}
\thanks{$^2$ is with the Institute of High Performance Computing (IHPC), Agency for Science, Technology and Research (A*STAR), 1 Fusionopolis Way, \#16-16 Connexis, Singapore 138632, Republic of Singapore. Email: mohasha\_sampath@ihpc.a-star.edu.sg}
\thanks{$^3$ are with the School of Electrical and Electronic Engineering, Nanyang Technological University, Singapore. Email: \{anshuman004; elkgoel; ehbgooi; hunghtd\}@ntu.edu.sg}
\thanks{$^\star$ Corresponding author: Hung Dinh Nguyen}
}

\maketitle

\begin{abstract}
This work proposes a novel degradation-infused energy portfolio allocation (DI-EPA) framework for enabling the participation of battery energy storage systems in multi-service electricity markets. The proposed framework attempts to address the challenge of including the \textit{rainflow} algorithm for cycle counting by directly developing a closed-form of marginal degradation as a function of dispatch decisions. Further, this closed-form degradation profile is embedded into an energy portfolio allocation (EPA) problem designed for making the optimal dispatch decisions for all the batteries together, in a shared economy manner. We term the entity taking these decisions as `facilitator' which works as a link between storage units and market operators. The proposed EPA formulation is quipped with a conditional-value-at-risk (CVaR)-based mechanism to bring risk-averseness against uncertainty in market prices. The proposed DI-EPA problem introduces fairness by dividing the profits into various units using the idea of marginal contribution. Simulation results regarding the accuracy of the closed-form of degradation, effectiveness of CVaR in handling uncertainty within the EPA problem, and fairness in the context of degradation awareness are discussed. Numerical results indicate that the DI-EPA framework improves the net profit of the storage units by considering the effect of degradation in optimal market participation.
\end{abstract}
\begin{IEEEkeywords}
Degradation-aware Dispatch, Energy Portfolio Allocation, Energy Storage, Conditional-Value-at-Risk. 
\vspace{-2mm}
\end{IEEEkeywords}

\section{Introduction}
\textcolor{black}{The electricity markets over the world are making regulatory changes in order to encourage participation of energy storage systems (ESS) \cite{bjorndal2023energy}. The ESS offer services essential for managing the intermittency associated with renewable energy and for lowering peak demand \cite{sayfutdinov2019degradation}. The ESS can perform energy arbitrage in electricity markets by charging at lower prices and discharging at higher prices. Thus, its revenue depends on the price gap which varies depending on the energy supply and demand in the network. Consequently, the revenue is subject to market risks due to price volatility. }

\textcolor{black}{An energy storage aggregator is a coalition of multiple energy storage systems (ESS) that collectively participates in electricity markets on their behalf. Forming such a coalition offers several advantages, including optimized resource utilization, enhanced market power, and effective risk hedging~\cite{diaz2022coordination}. Additionally, it enables smaller units to access electricity markets that they might otherwise be excluded from due to minimum battery size requirements.}
  
The market participation strategy of an ESS is essentially an optimization problem subject to the battery's operational and physical limits as constraints \cite{lund1, sayfutdinov2022optimal}. A large number of these dispatch studies do not consider degradation in the optimization problem \cite{singh2022two,dimitriadis2022strategic,lau2023stochastic}. \textcolor{black}{ According to some standards, once the battery reaches a 20\% loss in its capacity, it is deemed unfit \cite{andrenacci2021battery}. Therefore, storage systems with higher degradation rate or with a higher replacement cost may adjust the bidding strategies to account for the higher degradation costs \cite{mostafa2017long,maheshwari2020optimizing}. The degradation model helps optimize dispatch to balance revenue generation with long life.} Also, a degradation-aware dispatch can help achieve a circular economy with batteries \cite{9783055}.

There are multiple approaches to estimate the loss in capacity due to cycling. The stress factor-based capacity degradation models are widely used to estimate the incremental degradation due to battery cycling \cite{cordova2021energy}. The stress factor models are semi-empirical models that express the loss in battery capacity as a function of battery states such as the depth-of-discharge (DoD), temperature, state-of-charge (SoC), etc., in each equivalent charge-discharge cycle~\cite{YLI2020}. The equivalent cycles are obtained by the \textit{rainflow} cycle counting algorithm which utilizes the SoC profile traversed by the battery over a time horizon to extract the equivalent half or full cycles of charge and discharge \cite{xu2017factoring}. Particularly in a multi-market environment where the ESS goes through partial charge-discharge cycles, a cycle counting algorithm is needed for efficient estimation of capacity loss. \textcolor{black}{However, the rainflow algorithm does not possess an analytical functional form as it is a set of rule-based logical statements, and hence, it is challenging to incorporate it into an optimization framework.} Specially designed methods have been explored in \cite{lai2022profit,lee2022novel,shi2018optimal,anshuman_scobd} to incorporate cycle counting based storage degradation into battery dispatch frameworks. Authors in \cite{lai2022profit} have utilized rainflow algorithm-based battery degradation cost in the training process of its reinforcement learning-based ESS scheduling model. The authors in \cite{bansal2021storage} have proposed a graph-based linear mapping of SoC profile to charge-discharge half-cycle depths which exploits the rainflow algorithm. In \cite{lee2022novel}, a stress factor-based degradation model is proposed which counts degradation only during battery discharging and thus, removing the need for a cycle counting algorithm in the optimization framework. However, it may lead to over- or under-estimation of the incremental degradation. \textcolor{black}{The authors in \cite{shi2018optimal} have developed an online control policy for optimal battery energy dispatch problem considering the rainflow algorithm-based degradation cost. The control policy is developed assuming known penalty parameters for a certain market horizon. Further, \cite{shi2018optimal} develops a sub-gradient method based iterative procedure for solving the optimization problem where the optimality is guaranteed under some mild conditions \cite[Sec.~III-D and Theorem~3]{shi2018optimal}. However, none of the above works focus on developing a closed-form expression for estimating battery capacity degradation which can be \emph{directly} incorporated into the energy storage optimization problem and helps solving it in a single-level using the off-the-shelf solvers, without demanding parameter tuning or additional iterations.  
}

Another contribution of this work is the creation of a bidding strategy for the participation of a storage coalition in electricity markets. Various studies explore the integration of sharing economy frameworks into battery participation in markets~\cite{khojasteh2022robust,ritter2019sharing,chakraborty2018sharing,zhu2021bilevel}. A comprehensive review of different approaches to solving shared storage problems can be obtained from~\cite{dai2021utilization}. These works highlight the importance of coalition based approaches for market participation. However, the challenge of participation with \emph{volatile} market prices of future instances remains a challenge not addressed in these works. There have been attempts to employ the portfolio theory to solve the participation under uncertainty \cite{zheng2023bidding,han2022reducing,wang2020day}. Mean-variance optimization has also been used for storage scheduling~\cite{fang2018mean}.

Importantly, due to uncertainties in price information, the distribution of coalition profit maximization \cite{chakraborty2018sharing} does not follow a specific probability distribution like the normal distribution and is not symmetric. Therefore, the Markowitz model or mean-variance optimization \cite{fang2018mean,ma2023portfolios} fails to capture the skewness in the EPA objective distribution \cite{kisiala2015conditional}. Further, the variance is not monotonic, and thus, lacks the coherency property as a risk measure \cite{kisiala2015conditional}. These issues highlight the need to develop a shared economic participation framework for energy storage systems that can handle price uncertainties and incorporate degradation in dispatch decisions.

In this paper, we propose a degradation-infused energy portfolio allocation (DI-EPA) framework that provides optimal procurement of services from electricity markets, including energy, reserve, and regulation. Additionally, considering the complexities of existing degradation estimation methods, in this work, we express the incremental degradation as a function of dispatch and utilize the Gaussian process (GP)~\cite{williams2006gaussian} to learn its closed-form~\cite{pareek2021framework}, thus, eliminating the need for a cycle-counting algorithm. This closed-form expression of battery degradation is incorporated into the proposed DI-EPA framework that also considers the conditional value-at-risk (CVaR) approach to ensure a minimum profit guarantee under uncertain market clearing prices. This CVaR-based constraint mitigates the risk of ending a given scheduling period with financial loss. As such, this work approaches the DI-EPA problem from the optimization modelling's point of view. Moreover, we do not attempt to propose a new market model but provide a fair energy storage participation framework within existing market structures. The main contributions of this work are summarized below.
\begin{itemize}
    \item Proposing a novel risk-averse and degradation-infused energy storage coalition framework for participating in a multi-service electricity market based on a shared-economy model. The proposed framework improves the economic utilization of batteries over their lifetime by strategically dispatching them according to their degradation profiles.
    \item Developing a Gaussian Process-based \emph{degradation and dispatch} (D\&D) profiling to model the cost of degradation (COD) of each battery unit as a function of its dispatch. 
    This profiling infuses the degradation awareness into decision-making for shared-economy operations. 
    \item Formulating and solving an EPA problem with consideration of CVaR for market price and uncertainties associated with degradation profiling, aimed at introducing risk aversion into the bidding strategy. Additionally, we devise a fair mechanism for profit distribution by drawing upon concepts from marginal contributions within the Shapley value framework to equitably distribute overall profits.
\end{itemize}
 
\begin{remark}
The battery/storage participation via an aggregator has been studied in \cite{Contreras2019,chakraborty2018sharing}. The majority of these works, both cooperative and comparative participation, model the aggregator as an independent entity that has its own goals-- profit making for itself in general. Studies have also been carried out on how not to let the aggregator tilt the balance in its favor by altering prices \cite{Contreras2019}. However, in the proposed framework, the entity designed for market participation does not pursue its own profit making goals and only provides some `facilities' to the battery owners. These facilities or services include the participation of batteries in electricity markets and fair distribution of profit at the end of participation. In return, the battery operators can work on a fixed fee or profit share model to compensate for the service. Therefore, to distinguish the workings from a profit-seeking aggregator, we term the entity providing these facilities as `Facilitator'. The services provided by the facilitator can also include the remaining life estimation and second life prediction, etc.
\end{remark}
In the following, \textbf{we use small bold letters to indicate vectors of appropriate dimension}, e.g., $\mathbf{d}=[d_1,d_2,\ldots]^\top$ as dispatch vector; $\mathbf{1}$ is the vector of ones of appropriate dimension; and $N!$ represents the factorial of the whole number $N$.

\vspace{-2mm}
\section{Facilitator Framework}\label{sec:framework}
\begin{figure}[tb]
    \centering
    \includegraphics[width=\linewidth]{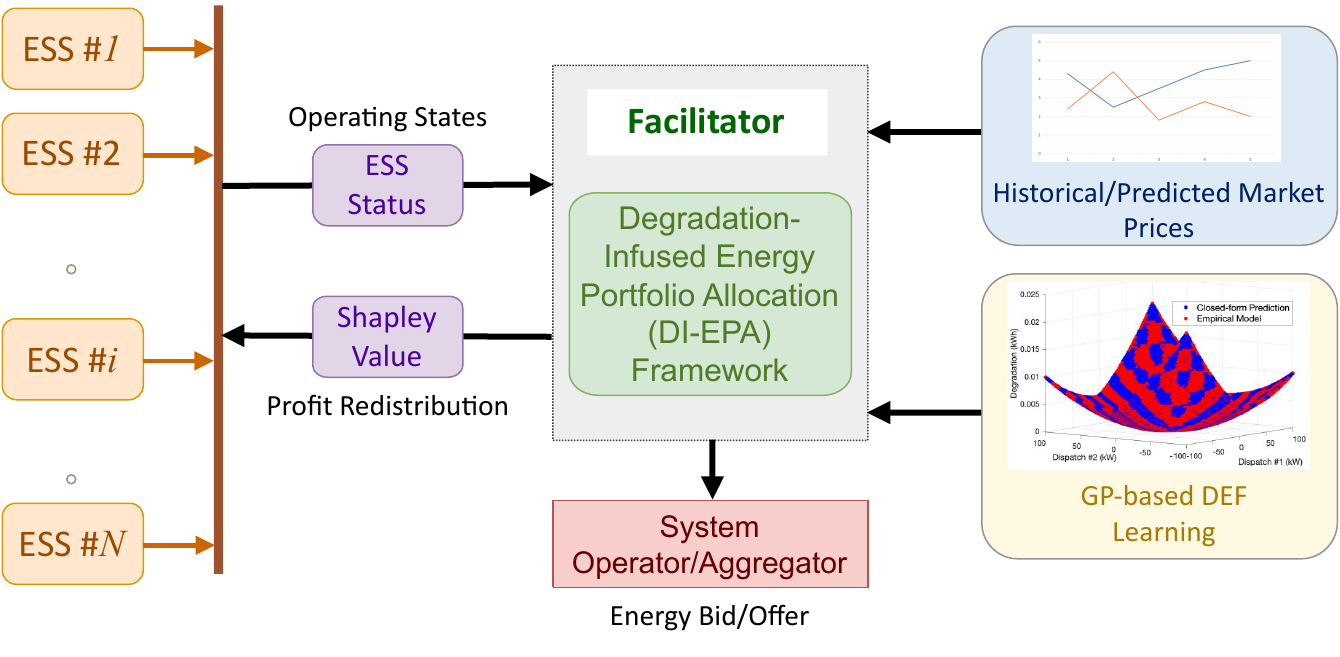}\vspace{-2mm}
    \caption{Structure of the proposed Facilitator framework}
    \vspace{-4mm}
    \label{fig:diepa}
\end{figure}

This section presents the core idea of the facilitator framework and defines the concept of degradation-infused fairness (DIF) for batteries operating under the sharing economy setting. We further formulate the problem of energy portfolio allocation (EPA), and discuss the features and merits of the proposed framework. To this end, Fig.~\ref{fig:diepa} illustrates the overall structure of the facilitator framework including its salient features.

The idea of DIF is built on the concept that economic payback from a battery relies heavily on the manner in which it is operated. This means that by modifying the operation of the battery, the total earnings over the life-cycle can be improved. For storage participation under the sharing economy setting, the concept of DIF has two implications. Firstly, to achieve better life-cycle payback on investment, the overall degradation of the storage coalition (or group) must be considered while participating in the market operations. The necessity of consideration of degradation gets aggravated if batteries procure multiple services from electricity markets simultaneously, including energy, reserve, and regulation. The underlying reason for this is that different services lead to different operational conditions for batteries which means a different level of degradation, even if the monetary payback is the same. The second implication of DIF in the sharing economic setting is that payback and degradation together constitute the net profit and must be distributed fairly. Further, a fair distribution of degradation will only depend on the operation (dispatch decisions) of a battery and thus will be linked with the market operating decisions taken for the storage coalition. We propose to call the entity making decisions on behalf of the storage coalition (group of storage units participating together in sharing economy settings) as the \emph{facilitator}.

\vspace{-2mm}
\subsection{EPA Framework}
The facilitator procures multiple services from electricity markets, via aggregators or market operators\footnote{We use market operators or aggregators interchangeably as facilitator can engage with any of them without the loss of generality.}, to earn revenue by providing various services supporting network/grid operations. 
Moreover, due to the temporal dependencies of decisions and states in the storage operation, facilitator participation is a multi-period/intertemporal optimization problem 
The locational marginal prices (LMPs) of future time instances, being forecasted values, introduce uncertainties in facilitator decisions, impacting its profit/welfare. Managing these uncertainties is crucial for the facilitator's efficient operation. To achieve these characteristics, we formulate the EPA model of the facilitator as

\color{black}
\begin{equation}\label{eq:faci}
\begin{aligned}
\hspace{-12mm}\max_{\substack{{\cal D}, {\cal X}}} \ & F(\Lambda,{\cal D}) = \sum^{}_{j \in {\cal B}} \bigg \{ \underbrace{\Tr\{\Lambda^\top D_j\}}_{\textsf{Revenue Term}} - \underbrace{c_jg_j(D_j)}_{\textsf{COD Term}} \bigg \}\\
\st \ & f^{\rm O}_j(D_j,\mathbf{x}_j) \leq 0; \quad \forall j \in {\cal B}\\
\ & f^{\rm M}_m({\cal D}) \leq 0;  \quad \forall m \in {\cal M} \\
& \underbrace{F_\alpha(\Lambda,\mathcal{D})}_{\textsf{\rm VaR}} \geq \varphi^\star.
\end{aligned}
\end{equation}

In \eqref{eq:faci}, $D_j = [\mathbf{d}^1_j, \ldots, \mathbf{d}^H_j]$ is the collection of dispatch decisions of the $j$-th ESS where $j \in {\cal B}=\{1,\ldots,B\}$. The vector $\mathbf{d}^h_j = [d^h_{j,1},\ldots,d^h_{j,M}]^\top$ is the dispatch decision vector at market interval $h \in {\cal H}= \{1,\ldots,H\}$ over a set of different markets/services ${\cal M}=\{1,\ldots,M\}$. Thus, $D_j$ is an $M\times H$ matrix, and ${\cal D}=\{D_1,\ldots,D_B\}$. Similarly, the market price matrix $\Lambda$ constitutes LMP values over all the markets and instances $\,\Lambda =[\boldsymbol{\lambda}^1, \ldots, \boldsymbol{\lambda}^h]$ where $\boldsymbol{\lambda}^h \in \mathbb{R}^{M}$. Also, symbol $\rm Tr$ in the objective represents the trace operator. 

\color{black}
The first term in the objective of \eqref{eq:faci} represents the revenue earned from participating in energy markets. The second term in the objective represents the COD as a function of dispatch by each ESS. The function $g_j(D_j)$ is the D\&D profile or the degradation estimation function (DEF) of the $j$-th ESS in the functional form while the COD coefficient $c_j$ is the cost multiplier (in \$/kWh) to represent the capacity degradation in terms of monetary cost. The most obvious choice of COD coefficient is the levelized COD-based multiplier \cite{7491274}. By combining the revenue and COD terms, the proposed EPA problem brings DIF within the sharing economy paradigm.

In the proposed EPA formulation~\eqref{eq:faci}, there are two different types of constraints that are technically relevant and financially interesting for the facilitator operation. Firstly, the function $f^{\rm O}_j$ represents the operational and capacity constraints of individual ESSs. 
Then, the function $f^{\rm M}_m$ represents market constraints such as the maximum power trading amounts that the storage coalition must abide by in each market. \textcolor{black}{These constraints are given in detail in the next section.}

The second constraint type (i.e. function $F_{\alpha}$) addresses the challenge of uncertainty in future market price/LMP predictions. 
Therein, we adopt VaR and CVaR concepts as risk measures, enabled by \cite{rockafellar2000optimization}. Specifically, VaR (and CVaR) are functions of all dispatch variables $\mathcal{D} = {D_1, \ldots, D_B}$ and are utilized to limit the risk of facilitator's profit loss below a predefined threshold $\varphi^\star$, with a given confidence level $\alpha$.

\vspace{-4mm}
\subsection{Constraints in the EPA Framework}
\color{black}
The constraints of the EPA  optimization model are developed in this section. Let $P^c_{h,j,m}$/ $P^d_{h,j,m}$ represent the charging/discharging power of ESS $j \in {\cal B}$ for market index $m \in {\cal M}$ and time index $h \in {\cal H}$, respectively. Further, let $\tilde{P}_{h,j}^c$/ $\tilde{P}_{h,j}^d$ be the power through ESS $j$ aggregated over all markets. Hence,
\begin{equation} \label{eq:m1}
    \begin{aligned}
            \tilde{P}^c_{h,j} = \sum_{m \in \mathcal{M}} P^c_{h,j,m} ; \ \
    \tilde{P}^d_{h,j} = \sum_{m \in \mathcal{M}} P^d_{h,j,m}
    \end{aligned}
\end{equation}
Now, the state of charge ($\xi$) for each ESS can be estimated as
\begin{equation} \label{eq:m2}
\begin{aligned}
        \xi_{h,j} & = \xi_{h-1,j} - \big( \frac{\tilde{P}_{h,j}^d}{\eta^d_j} - \eta^c_j \, \tilde{P}_{h,j}^c \big) \, \frac{\Delta_t}{E_j} \\
        0 & \leq \zeta_{h,j} \leq 1 \quad \forall \ h \in \mathcal{H} , j \in \mathcal{B} 
\end{aligned}
\end{equation}
Here, $\eta^c_j$/ $\eta^d_j$ represents the charge/discharge efficiency of ESS $j$, respectively. $\Delta_t$ represents the size of the dispatch interval in hours. $E_j$ represents the rated capacity of ESS $j$ in kWh.

To avoid simultaneous charge-discharge results of the battery, the operational limit on dispatch is modeled in terms of binary variables as follows
\begin{equation} 
    \begin{aligned}
        0 &\leq P^c_{h,j,m} \leq \beta^c_{h,j,m} \overline{P}_j^c \\
        0 &\leq P^d_{h,j,m} \leq \beta^d_{h,j,m} \overline{P}_j^d  \\
        & \beta^c_{h,j,m}  + \beta^d_{h,j,m} \leq 1
    \end{aligned}
\end{equation}
where $\beta^c_{h,j,m}$/$\beta^d_{h,j,m}$ represents the binary variables corresponding to charging and discharging variables. $\overline{P}_j^c$/ $\overline{P}_j^d$ represents the operational limits on charge/discharge power for $j^{th}$ ESS. 
Finally, the charge-discharge power can be linked to dispatch variable $d$ used in \eqref{eq:faci} as 
\begin{equation}
    \begin{aligned}
        d_{j,m}^h = P^c_{h,j,m} + P^d_{h,j,m} 
    \end{aligned}
\end{equation}

The maximum power trading limits (on the Facilitator) for any particular market/service $m$ can be modeled as follows
\begin{equation}\label{eq:m3}
    \begin{aligned}
        \sum_{j \in \mathcal{B}} P^d_{h,j,m}  \leq tl_m , \quad
        \sum_{j \in \mathcal{B}} P^c_{h,j,m}  \leq tl_m 
    \end{aligned}
\end{equation}
where $tl_m$ is the trading limit for market $m$.

\subsection{Fair profit-sharing mechanism}
\textcolor{black}{In this work, ESS units cooperate to participate in various markets like energy and reserve, where both prices and risks differ. Each ESS prefers markets with better prices and lower risks, but must cooperate due to trade limits. Proportional distribution can under-compensate ESSs in lower-priced markets or higher-risk markets, despite their crucial role. The Shapley value pools profits from all markets and allocates them based on each ESS’s marginal contribution, ensuring a fairer profit share for those that reduce the aggregator’s risk. In the later sections, we develop the detailed mathematical model for shapley value (refer to \eqref{eq:shaply}).}

\color{black}
\section{D\&D Profiling and COD Formulation}\label{sec:DnD}
This section describes the motivation behind opting to develop a D\&D profile and the requirements for obtaining such a profile, followed by the GP-based procedure to learn a closed-form degradation profile. 
 
The battery degradation is estimated using the cycle counts provided by the \textit{`rainflow algorithm'}-- as in{ \cite{mostafa2017long,maheshwari2020optimizing,murli2018coordinated} -- which counts the equivalent cycles for a given SoC profile. However, the inclusion of rainflow into an optimization problem is challenging as it is in the form of nested \textit{if-else} conditional statements. 

A linear approximation-based method has been proposed in \cite{xu2017factoring}. Yet, the limitation lies in the fact that the stress function is near quadratic in nature and rainflow is only used for bench-marking, not within the optimization problem.
Further, consideration of degradation is more challenging when the motive is to obtain a fair-degradation profile while working with multiple batteries simultaneously as in the case of the facilitator's DI-EPA problem \eqref{eq:faci}. An idea to solve this problem can be to develop an `optimizable' functional representation of the battery degradation. This representation should not only be able to approximate the battery degradation for a given dispatch vector but also should be solvable within the market participation or EPA problem. Mathematically, we need a function that provides an approximate degradation value given a dispatch vector. This is because, in the DI-EPA \eqref{eq:faci}, the dispatch is the decision vector and should be able to control both the revenue and COD. Further, the DEF required in \eqref{eq:faci} must be differentiable in a closed-form. This is important because simulation models or complex data-driven models  (e.g., deep neural networks) are typically `optimizable' within a market participation problem. Another imperative property of the DEF $g(\mathbf{d})$ must be to disconnect the degradation estimation process from the `rainflow' algorithm-based cycle estimation. We also want an interpretable and reproducible model for estimating the degradation. As the GP belongs to the Bayesian paradigm, we only need to provide information about training data and kernel choice for the model to be reproducible \cite{williams2006gaussian}. Further, a closed-form can be obtained upon learning \cite{pareek2020gaussian,pareek2021framework}. Moreover, GP can help to learn the direct function between degradation and dispatch, without requiring cycle computations. Therefore, we employ the GP as a regression tool to obtain $g(\mathbf{d})$.

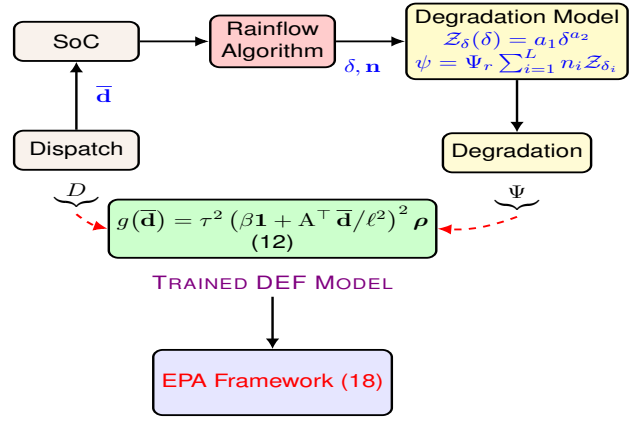
\begin{figure}[t!]
	\centering
	\resizebox{0.95\columnwidth}{5.5cm}{
		\tikzset{every picture/.style={line width=1pt}}
		\begin{tikzpicture}[node distance=1.5cm,
			every node/.style={fill=white, font=\sffamily}, align=center]
			\node (ip)[rectangle, rounded corners, fill = brown!10,draw = black, minimum width = 1.8cm,minimum height = 0.75cm] {{Dispatch}};
			\node (ip2)[above = 1.3cm of ip,rectangle, rounded corners, fill = brown!10,draw = black, minimum width = 1.8cm,minimum height = 0.75cm] {{SoC}};
			\node (ip3)[right = 1cm of ip2,rectangle, rounded corners, fill = red!20,draw = black, minimum width = 1.8cm,minimum height = 0.75cm]{Rainflow \\ Algorithm};
			\node (model) [right = 1cm of ip3 , rectangle, rounded corners, draw = black , fill=yellow!25, minimum width=1.5cm,minimum height=1cm]{Degradation Model\\ \textcolor{blue}{ ${\cal Z}_{\delta}(\delta)  = a_1 \delta^{a_2}$} \\ \textcolor{blue}{$  \psi = \Psi_r \sum_{i=1}^L n_i {\cal Z}_{\delta_i} $}};
			\node (op) [below = 1cm of model, rectangle, rounded corners, ,draw = black,fill = yellow!20,minimum width = 1.8cm,minimum height = 0.75cm] {{{Degradation}}};
			\draw[->](ip3)  -- (model);
			\draw[->](ip)  -- (ip2);
			\draw[->](ip2)  -- (ip3);
			\node (a)[ above = 0.4cm of ip, xshift= 0.4 cm,fill = none]  {\textcolor{blue}{$\overline{\mathbf{d}}$}};
			\node (a)[ below = 0.1cm of ip, fill = none]  {$\underbrace{D}$};
			\node (aa)[ below = -0.3cm of ip3, xshift= 1.3cm, fill = none]  {\textcolor{blue}{$\delta,\mathbf{n}$}};
			\draw[->](model)  -- (op);
			\node(b)[below = 0.1cm of op, fill = none]  {$\underbrace{\Psi}$};
			\node(proxy)[below = 2.5cm of ip3,rectangle, rounded corners, ,draw = black,fill = green!20,minimum width = 1.8cm,minimum height = 1cm]{ $g\big({\overline{\mathbf{d}}}\big) = \tau ^2 \left (\beta \mathbf{1}+{\rm A^\top\,\overline{\mathbf{d}}}\big / {\ell ^2} \right)^2 \boldsymbol{\rho} $ \\ (12) };
			\draw[->,red,dashed] (a.south) to[bend right=15] (proxy.west);
			\draw[->,red,dashed] (b.south) to[bend left=15] (proxy.east);
			\node(c)[below = 0.1cm of proxy, fill = none, yshift = -0.1cm]{\textcolor{violet}{\textsc{Trained DEF Model}}};
			\node (epa)[below = 1cm of c,rectangle, rounded corners, fill = blue!10,draw = black, minimum width = 2.8cm,minimum height = 1.25cm] {\textcolor{red}{EPA Framework  (18)}};
			\draw[->](c)  -- (epa);
	\end{tikzpicture}}
	\vspace{-1mm}
	\caption{The degradation estimation function learning mechanism along with the degradation model used.}
	\label{fig:learning}\vspace{-1mm}
\end{figure}

\vspace{-4mm}
\subsection{Degradation Model and Dataset Generation} \label{sec:deg_model}
Fig. \ref{fig:learning} shows the process of DEF learning along with the degradation model used in this work. The goal of the proposed DEF learning model is to estimate the incremental loss in battery capacity for a given sequence of charge-discharge decisions. The input to the learning model $\overline{\mathbf{d}}$ represents the total dispatch by one battery over the market horizon, obtained by summing the market-wise dispatch.
\begin{equation}
    \overline{\mathbf{d}} = {\mathbf{1}^\top D =} [\mathbf{1}^\top\mathbf{d}^1, \ldots, \mathbf{1}^\top\mathbf{d}^H]
\end{equation}
Here, index $j$ corresponding to the battery units is omitted for brevity. A dispatch sample $D$ is passed through the SoC relationship function which provides the SoC level at each step based on $\overline{\mathbf{d}}$ and the initial SoC. The next step in the learning degradation is to quantify the charge-discharge cycles into equivalent full or half cycles. This is achieved by providing the samples that passed the SoC feasibility test\footnote{Here, SoC feasibility test means that if a dispatch vector takes the battery SoC below (above) the minimum (maximum) permissible SoC levels (say 20\% and 80\%), then such a dispatch vector is considered infeasible.} to the rainflow cycle counting algorithm which in turn calculates the number of equivalent cycles and the starting-ending points for each cycle. Full details of the rainflow algorithm can be found in \cite{tsaha}. With this information, the stress on battery ${\cal Z}$ in each equivalent cycle is estimated as a function of DoD as follows:
\begin{align}\label{eq:stress}
     {\cal Z}^{}_{\delta}(\delta)  = a_1 \delta^{a_2}  
\end{align}
where, $\delta$ is the DoD of the corresponding cycle, and $a_1$ and $a_2$ are model coefficients. The cumulative stress based capacity degradation function is expressed as
\begin{align}\label{eq:deg}
       \psi = \Psi_r \sum_{i=1}^L n_i {\cal Z}^{}_{\delta_i} 
\end{align}
where $\mathbf{n}$ is a vector of length $L$ obtained from the rainflow algorithm that quantifies each equivalent cycle as full ($n_i=1$) or half cycle ($n_i=0.5$). $L$ is the total number of equivalent cycles in the given sample $D$, and $\Psi_r$ is the rated capacity of the battery. The degradation $\psi$ is given in kWh. 

Although it is straightforward to understand that using a large number of samples of $D$, one can learn the relationship between degradation and dispatch via a regression tool, there are two considerable challenges to overcome. The first challenge is to generate the training data such that it maps the large hypercube, formed by the multi-period feasible dispatch space of $D$. The second challenge is to provide the closed-form of DEF upon learning. To this end, we first present a method to sample in large dimension space for training data \cite{pareek2021framework} and later, present the GP-based closed-form DEF learning. 

The target of the D\&D profiling is to learn a degradation function with multi-period dispatch as the input. This means that DEF must be valid within a given dispatch hypercube (or subspace). The length of dispatch vector $H$ decides the hypercube dimension and it is constructed respecting the physical constraints of charging and discharging limits. Developing the training and testing datasets in this high-dimensional hypercube is challenging due to the \textit{curse of dimensionality}. Particularly, for DEF to have similar accuracy within a complete $H$-dimensional hypercube, the learning dataset must map the hypercube appropriately. The construction of a mesh-grid to obtain an evenly distributed set of samples is not feasible as the total number of points for a grid having $y$ number of points in each dimension grows with $y^{D}$ \cite{pareek2021framework} for each time instance, which grows even faster with $H$. It shows exponential growth even with modest values of $y$. A point to note here is that any aggregated dispatch vector $\overline{\mathbf{d}}$ must have each time instance entry within the charging limit of that battery. Also, not all such vectors might be feasible for SoC constraints, depending on the initial SoC. For example, if the initial SoC is at the maximum limit of SoC, we cannot charge the battery at the first instance, even if dispatch is within the charging limits. However, the SoC limits are always restrictive in nature, and thus, if a DEF works over a hypercube constructed using charging limits, it will always work with additional SoC constraints. Also, to be able to map the entire space within the charging limits, we use the dataset generation method described in \cite{pareek2021framework}.

Next, we deal with the challenge of providing a closed-from of DEF using GP learning. The subsection starts with the GP learning overview and leads to the DEF learning. 

\vspace{-2mm}
\subsection{GP-based DEF Learning}
The GP is a function approximation tool which is also termed as regression over function \cite{williams2006gaussian}.  
To this end, we generate total $N$ samples of total dispatch\footnote{Generating samples of $D$ and summing them market-wise is equivalent to directly generating samples of the aggregated dispatch $\overline{\mathbf{d}}$. This is due to the physical reality that the battery only experiences the total dispatch.} and obtain $\{\overline{\mathbf{d}}^1, \dots, \overline{\mathbf{d}}^N\}$,  following the procedure explained in Section~\ref{sec:deg_model}. The corresponding degradation is obtained using the mechanism described in Fig. \ref{fig:learning} as $\left\{\psi^1, \dots, \psi^N\right\}$. We collect all these training data samples as $\{\rm A,\boldsymbol{\psi}\}$ where ${\rm A}$ is the \textit{design matrix} of size $ N \times H$ with all the aggregate dispatch samples $\overline{\mathbf{d}}$ collected as rows while $\boldsymbol{\psi}$ is a vector having $N$ degradation values. Now, one of the most significant features of GP is to allow prior knowledge integration into the learning model. A promising way for doing this is by selecting a suitable Kernel or covariance function $k(\cdot,\cdot)$ \cite{williams2006gaussian,pareek2021framework}. This Kernel or covariance function establishes the input-output relationship using the idea that if two inputs are `nearby', then the corresponding two outputs will also be close to each other. 

In selecting an appropriate kernel function, we use the information regarding the properties of the degradation function \eqref{eq:deg}. Firstly, it can be observed from \eqref{eq:stress} that stress is a polynomial function of the DoD. Further, the degradation function \eqref{eq:deg} can be interpreted as a weighted sum of polynomial functions \eqref{eq:stress}. Another property of \eqref{eq:deg} can be stated that it is a smooth function. Further, as DoD is linearly related to SoC and power dispatches, SoC is also obtained by linear operations on the dispatch vector. Hence, the degradation $\psi$ has a smooth polynomial relationship with aggregated dispatch $\overline{\mathbf{d}}$. The degree of such a polynomial relationship directly depends on the value of $a_2$ in~\eqref{eq:stress}. A well-expected degradation model for Li-Ion cells \cite{xu2017factoring} is quadratic in nature with $a_2 = 2.03$. Thus, we opt for the polynomial kernel of degree two as
\begin{align}\label{eq:poly}
     k(\rm A,\overline{\mathbf{d}}) & = \tau ^2 \left (\beta \mathbf{1}+{\rm A^\top\,\overline{\mathbf{d}}}\big / {\ell ^2} \right)^2
\end{align}
Now, using \eqref{eq:poly}, a GP model obtains the mean estimate of the degradation $\psi$ \eqref{eq:deg}, as a function of dispatch, using \textit{maximum a posteriori} estimation.  Let $g(\cdot)$ represent the degradation function as an estimate shown in Fig. \ref{fig:learning}. Then, the mean prediction of the degradation can be obtained as
\begin{align}  
    g(\overline{\mathbf{d}}) = \mathbf{k}^\top(\rm A,\overline{\mathbf{d}}) \big [k(\rm A,\rm A)+\sigma^2_\epsilon I \big]^{-1}\boldsymbol \psi \label{eq:deg_gp}
\end{align}
\color{black}
Finally, we express \eqref{eq:deg_gp} in compact form as
\begin{align}
    g(\overline{\mathbf{d}})  & = \tau ^2 \left (\beta \mathbf{1}+{\rm A^\top\,\overline{\mathbf{d}}}\big / {\ell ^2} \right)^2\, \boldsymbol{\rho} \label{eq:deg_final}
\end{align}
where, $\boldsymbol{\rho}= \big [k(\rm A, \rm P)+\sigma^2_\epsilon \rm I \big]^{-1}\boldsymbol \psi$ , is constant upon learning and $\mathbf{k}$ indicates the kernel function of variable $\overline{\mathbf{d}}$ with the \emph{Design Matrix} $\rm P$. Equation \eqref{eq:deg_final} is the resulting degradation model and will be used in the proposed DI-EPA framework later.

\color{black}
From \eqref{eq:poly} and \eqref{eq:deg_gp}, it is clear that for each battery, the function $g(\overline{\mathbf{d}})$ will depend on a set of \textit{hyper-parameters} $\{\tau,\beta, \ell\}$. These \textit{hyper-parameters} numerically calculate the quadratic covariance (or kernel) function \eqref{eq:poly} upon learning with dataset $\{\rm A,\boldsymbol \psi\}$. Additionally, as $\sigma_\epsilon$ is included in $\boldsymbol{\alpha}$, GP can handle noisy data. A particular case for noisy data in this context is the errors involved in the empirical models of stress \eqref{eq:stress} and degradation \eqref{eq:deg}. In this manner, the proposed GP-based DEF learning captures the model errors, enhancing the flexibility of the framework. The structures of kernel matrix $K(\rm P,\rm P)$, kernel vector $\mathbf{k}$, and other GP-learning details can be obtained from \cite{pareek2021framework,williams2006gaussian}.
 
\begin{remark}
To obtain the COD coefficient $c_j$ in \eqref{eq:faci}, various methods, including replacement cost \cite{7491274} and the marginal COD \cite{he2020power}, can be employed. Unlike the marginal COD, which considers the battery operates in the same market and with the same objective function throughout its period of operation \cite{he2020power}, the proposed framework allows for multi-market participation. This flexibility ensures that the COD coefficient $c_j$ reflects the cost of the battery in the context of diverse market operations. Since all battery units determine the COD coefficient using the same principle, the fairness of the approach will remain unaffected.
\end{remark}

\section{EPA Framework with VaR/CVaR Co-Optimization}\label{sec:cvar}

The profit $F(\Lambda,{\cal D})$ earned by the storage facilitator by energy arbitrage is the difference between its revenue earned and cost paid from and to the market operator for energy sales and purchases during different time intervals respectively after deducting the COD. In the presence of uncertain market clearing prices $\Lambda$, each aggregator tries to optimize its bids $D_j$ corresponding to each storage such that a balance between the expected profit $F(\Lambda,{\cal D})$ and the involved risk is maintained. VaR is a useful metric for the assessment of the risk associated with the market revenue in a probabilistic manner for a given portfolio~\cite{AXUAN2021}. It indicates the minimum guaranteed profit attainable under a given confidence level. Here, ${\rm VaR}_\alpha$ denoted by $\zeta_{\alpha}$ can be defined as the \emph{maximum} profit value such that the probability of profit being lower than or equal to this value $\zeta_{\alpha}$ is $1-\alpha$ where $\alpha \in (0,1)$ is the confidence level. Mathematically, it is modeled as 
\begin{gather}
   {\rm VaR}_{\alpha}(F): \ \zeta_{\alpha} = \max\big\{z \, \big| \, \mathbb{P}(F(\Lambda,{\cal D}) \leq z) \leq 1-\alpha \big\}
\end{gather}
As only a few probability distributions offer exact deterministic reformulations, it may not be practical to represent $F(\Lambda,D_j)$ as a ${\rm VaR}_{\alpha}$ that guarantees a \emph{minimum} profit amount~\cite{Roveto_9099127}. In addition, it may not offer a convex (and differentiable) ${\rm VaR}_{\alpha}(F)$ formulation and hence may not be tractable to obtain the required profit optimality guarantees via off-the-shelf solvers. Alternatively, CVaR formulations presented in \cite{rockafellar2000optimization} allow a convex representation, wherein CVaR provides an estimated mean of the profit values which are lesser than the VaR value for the same confidence level used to estimate the VaR. For instance, the ${\rm CVaR}_{0.95}$ is the average of the expected profit for plausible profit values lesser than the ${\rm VaR}_{0.95}$ threshold~\cite{Roveto_9099127}, i.e., the mean of the 0.95-negative tail distribution of $F(\Lambda,{\cal D})$ as defined below.
\begin{gather} \label{eq:cvar_1}
    {\rm CVaR}_{\alpha}(F): \ \phi_{\alpha} = \mathbb{E}\big\{F(\Lambda,{\cal D}) \, | \, F(\Lambda,{\cal D}) \leq \zeta_{\alpha}\big\}
\end{gather}
where $\mathbb{E}$ is the expectation operator and ${\rm CVaR}_{\alpha} \leq {\rm VaR}_{\alpha}$. Here, CVaR is a better alternative to VaR as it is a coherent risk measure, positively homogeneous and convex. Assuming $f_{\Lambda}$ be the probability density function of uncertain market clearing prices $\Lambda$, \eqref{eq:cvar_1} can be reinterpreted as follows:
\begin{align}
{\rm CVaR}_{\alpha}(F): \ \phi_{\alpha} = \frac{1}{1-\alpha} \int_{F(\Lambda,{\cal D}) \leq \zeta_{\alpha}} \hspace{-5mm} F(\Lambda,{\cal D}) f_{\Lambda} d\Lambda
\end{align}

Let $\varphi^\star$ be the \emph{minimum expected profit target} of the storage facilitator. ${\rm CVaR}_{\alpha}$ provides a probabilistic bound for $F(\Lambda,{\cal D})$ over $\varphi^\star$ as well as a sense of loss if $F(\Lambda,{\cal D})$ falls below $\varphi^\star$. However, replacing ${\rm VaR}_{\alpha}$ with ${\rm CVaR}_{\alpha}$ might end up over-ensuring the profit targets resulting in an infeasible solution. Hence, along with ${\rm CVaR}_{\alpha}$ maximization, we also aim at keeping $\zeta_{\alpha}$ as close as possible to $\varphi^\star$ while ensuring it will not fall below. As such, it is a simultaneous attempt to control VaR and maximize CVaR, i.e., co-optimizing both quantities. Therewith, the storage facilitator targets a multi-objective optimization problem that jointly maximizes the \emph{minimum guaranteed profit} (or the \emph{robust profit}) while minimizing the risk associated with it subjected to a predefined confidence level $\alpha$. Thus, the following reformulation including ${\rm CVaR}_{\alpha}$ is expressed as follows \cite{Roveto_9099127}:
\begin{align}\label{eq:joint_opt}
{\cal O}_{\alpha}:\ \max_{\zeta \in \mathbb{R}} \ \zeta + \frac{1}{1-\alpha} \int_{\zeta \in \mathbb{R}} \hspace{-0mm} \big[ F(\Lambda,{\cal D}) -\zeta \big]^{-} f^{}_{\Lambda} d\Lambda
\end{align}
where $ x^{-} :=\min\{0,x\} $. Here, the confidence interval $\alpha$ influences the trade-off between the profit and the risk associated with it. Therefore, an optimized storage bidding strategy with higher $\alpha$ leads to a lower expected profit which is less risky and vice versa. A major limitation in formulation \eqref{eq:joint_opt} is that it requires the probability distribution/density function $f^{}_{\Lambda}$ of market clearing prices $\Lambda$. However, exact $f^{}_{\Lambda}$ or a close approximation to it is not readily available generally in practice. Hence, we modify the above formulation to infer the variation of market clearing prices $\Lambda$ in terms of a finite set of historical samples $\Lambda_{s}$ for all $s\in {\cal S}$ as follows.
\begin{align}\label{eq:joint_opt_sample}
\tilde{\cal O}_{\alpha}:\ \max_{\zeta \in \mathbb{R}} \ \zeta + \frac{1}{(1-\alpha)|\cal S|} \sum_{s \in {\cal S}} \big[ F(\Lambda_{s},{\cal D}) -\zeta \big]^{-}
\end{align}

\vspace{2mm}
\color{black}
Therewith, the overall storage DI-EPA optimization problem that determines the Facilitator's bidding strategy for a given market horizon is formulated as follows:
\vspace{2mm}
\begin{subequations}\label{eq:overall_di_epa}
\begin{align}
\tilde{\cal O}^{\star}_{\alpha}: \max_{\substack{{\cal D},\, \zeta \in \mathbb{R}  \\ \boldsymbol{\ell},\, \boldsymbol{\Gamma},\, {\cal X} }} \ \hspace{-8mm} & \hspace{8mm} \zeta +\frac{1}{(1-\alpha)|\cal S|} \sum_{s \in {\cal S}} \ell_{s} \label{eq:overall_di_epa_a} &  \\
\st \quad & \ell_{s} \leq \sum^{}_{j \in {\cal B}} \Big \{ \Tr\{\Lambda_s^\top {\cal D}\} - \Gamma_j \Big \} -\zeta;\ &\forall s \in {\cal S}  \label{eq:overall_di_epa_b} \\
& \ell_{s} \leq 0;\ & \forall s \in {\cal S}   \label{eq:overall_di_epa_c} \\
& \Gamma_j \geq c_j \tau ^2 \left (\beta \mathbf{1}+{\rm A^\top\,\overline{\mathbf{d}_j}}\big / {\ell ^2} \right)^2\, \boldsymbol{\rho} ;\quad & \forall j \in {\cal B} \label{eq:overall_di_epa_d} \\
& \zeta \geq \varphi^{\star} &  \label{eq:overall_di_epa_g} \\
& \eqref{eq:m1} - \eqref{eq:m3} &
\end{align}
\end{subequations}
\color{black}
where $\ell_s; \forall s \in {\cal S}$ is an auxiliary variable introduced for each price sample in the objective \eqref{eq:overall_di_epa_a}. $\ell_s$ with the constraints \eqref{eq:overall_di_epa_b}-\eqref{eq:overall_di_epa_c} jointly convert the \emph{infimum} part in the objective \eqref{eq:overall_di_epa_a}, i.e., the second term of \eqref{eq:joint_opt_sample}, to a simplified readily-usable form compatible with  off-the-shelf solvers, like GUROBI. $\Gamma_j; \forall j \in {\cal B}$ is another auxiliary variable used to capture the COD $c_j g_j(D_j)$ for each ESS. For the \emph{convex-quadratic} form of the COD function obtained in \eqref{eq:deg_final}, constraint \eqref{eq:overall_di_epa_d} can be interpreted as a second-order cone constraint that is tractable and scalable in solving. In case if the DI-EPA framework is to be used with different COD expressions, such as highly non-linear forms combined with convex approximation/reformulations, \eqref{eq:overall_di_epa_d} can be modified accordingly. 
Finally, VaR $\zeta$ guarantees the minimum profit target $\varphi^{\star}$ of the facilitator in \eqref{eq:overall_di_epa_g}.

Variable $\zeta$ in \eqref{eq:overall_di_epa} seeks the $\alpha$ confident minimum profit (or ${\rm VaR}_{0.95}$ and ${\rm CVaR}_{0.95}$ values of the profit) for the facilitator and does not represent how to allocate the profit individually for each ESS. Here, we use the \emph{Shapley value}\cite{shapley1971cores} concept to decide the profit rewarded to each ESS that accounts for the contributions made by each individual to the coalition. Accordingly, \eqref{eq:shaply} derives the marginal profit of ESS~$j$ considering its contributions under several possible storage coalitions ${\cal G} \subseteq {\cal B} $
\begin{equation}\label{eq:shaply}
\gamma_j = \hspace{-1mm} \sum_{{\cal G} \subseteq {\cal B}\setminus \{j\}} \hspace{-1mm} \frac{|{\cal G}|!(|{\cal B}|-|{\cal G}|-1)!}{|{\cal B}|!} \left[ \frac{ v_{{\cal G}\cup\{j\}} - v_{\cal G} }{v_{\cal B}}\right]
\end{equation}
where $v_{\cal B}$ is the VaR (or the minimum expected profit) created by the grand coalition, i.e., the group of all the ESSs ${\cal B}$; $v_{{\cal G}}$ is the VaR created by a given coalition ${\cal G} \subseteq {\cal B}\setminus \{j\}$ that is formed without the participation of ESS~$j$. Then, $v_{{\cal G}\cup\{j\}} - v_{\cal G}$ is considered as the contribution made by ESS~$j$ for a given coalition ${\cal G}\cup\{j\}$. As in \eqref{eq:shaply}, $\gamma_j$ computes the proportion of the profit that should be allocated to ESS~$j$ considering its overall contribution or its added value to the profit of the grand coalition by averaging out the contributions of ESS~$j$ over possible coalitions (permutations of participant ESSs).

\vspace{-2mm}
\section{Results and Discussion}\label{sec:results}
This section presents the simulation results and discussion on the proposed DI-EPA problem formulation and its components. First, we present a performance analysis of the D\&D profiling mechanism and then present a detailed case study on multi-market participation via DI-EPA. 
Fig. \ref{fig:DEF_2hr} represents the performance of the closed-form degradation function for two-instances\footnote{Here instance can be separated by any time interval say 15 min or an hour. As the degradation is only a function of dispatch, the time difference between instances has no effect on the D\&D profile performance.}. The marginal degradation empirical model represented in red in Fig. \ref{fig:DEF_2hr} is obtained after calculating cycles using the rainflow algorithm as described in Fig. \ref{fig:learning}. The shapes of both empirical and closed-form predictions closely resemble convex manifolds (bowl-shaped) which is due to the fact that the empirical model is close to quadratic in nature. This also validates the reason behind selecting the quadratic kernel \eqref{eq:poly}. The patterns of black and red dots represent the learning samples directly, and the closed-form can be tuned to achieve the desired accuracy. 

\begin{figure}[t!]
    \centering
    \includegraphics[width=0.8\columnwidth]{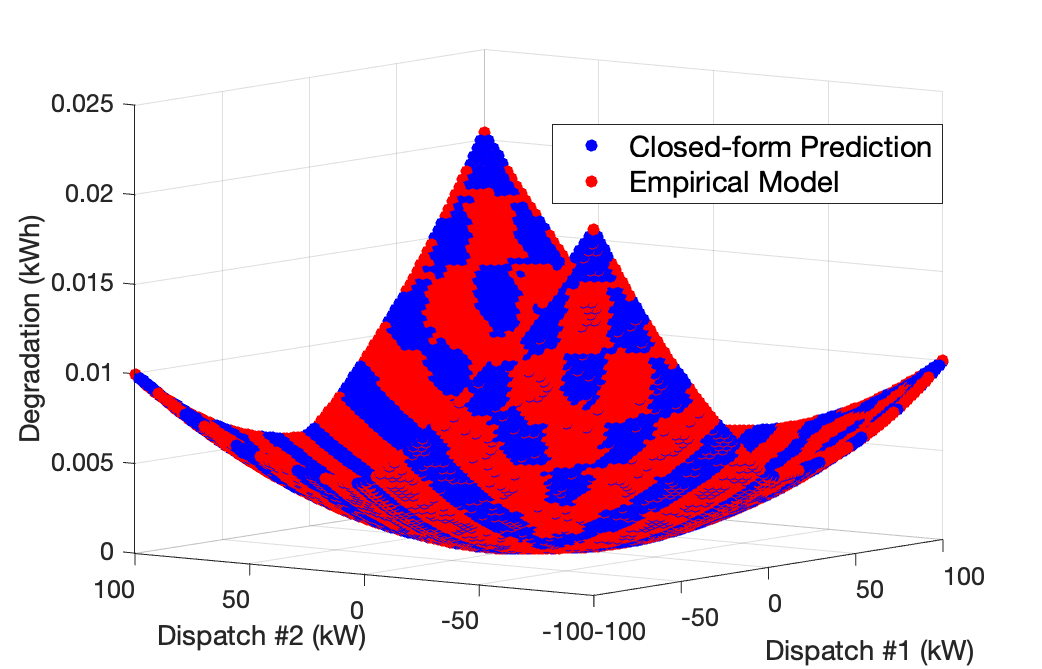}\vspace{-1mm}
    \caption{Empirical and closed-form prediction of degradation as a function of two-hour dispatch vector.}\vspace{-2mm}
    \label{fig:DEF_2hr}
\end{figure}

Further, to showcase the DEF learning performance on longer dispatch vectors ($H> 2$), Fig.~\ref{fig:deg_error} represents the cumulative distribution function (CDF) of error values for different time horizon ($H$) values. Here, the error calculation is done for a 500-kWh battery considering 15-min dispatch instances. The testing is carried out with $50000$ dispatch vectors $\overline{\mathbf{d}}$ of different lengths $H$. The CDF depicts that the degradation prediction error of the majority of samples lies within the $\pm 0.01$ kWh error margin for $H \leq 8$. This shows that the proposed DEF estimation can estimate the marginal degradation without depending on the rainflow cycle-counting algorithm. 
Next, we present the discussion on the DI-EPA problem solution for determining the dispatch vectors for multiple batteries procuring various services from electricity markets.

\begin{figure}[t!]
    \centering
    \includegraphics[width = 0.9\columnwidth]{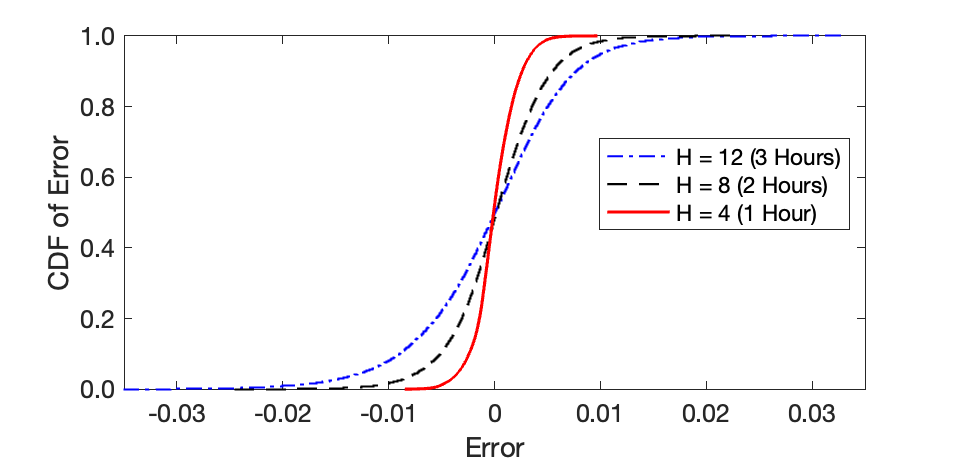}\vspace{-2mm}
    \caption{Empirical cumulative distribution function of Error in degradation estimation for a 500-kWh battery for different lengths of dispatch vectors.}\vspace{-2mm}
    \label{fig:deg_error}
\end{figure}

\vspace{-4mm}
\subsection{Multi-Market Participation}
Here, we conduct a case study where a facilitator hosts three battery ESSs that participate in both the energy and reserve markets. The market interval is taken as 15 minutes each for the two markets. In the energy market, the facilitator gets compensated according to its energy injection. In the reserve market, the facilitator first gets compensated for the capacity it commits for up or down reserve. Later, the facilitator gets paid or billed depending on the amount of energy discharged or charged respectively by the reserve market operator within the committed capacity. 

The details of the three battery ESSs are given in Table~\ref{table:data}. The three ESSs have equal power and energy ratings. The replacement cost for the three batteries are $\$$\{350, 450, and 550\}$\rm /kWh$. Based on the historical patterns, the facilitator uses a total of 250 price samples each for energy and reserve markets to optimize the dispatch schedule for all the batteries. For the case study, random samples were generated considering a 50$\%$ fluctuation from the mean price. The mean values of the price samples are shown in Fig.~\ref{fig:rate} for both the energy and reserve markets. \textcolor{black}{First, the proposed approach was compared with a risk-neutral approach by performing scheduling for $4$ market intervals. Later, the effect of confidence level $\alpha$ was analyzed on the expected profit as well as on the profit under risk.} Finally, the simulations were performed for a period of $40$ market intervals to analyze the effect of degradation cost on the net expected profit. Further for simplicity, it was assumed that the prices for up and down reserve allocations are equal. The simulations were performed on a computer with Intel$^{\circledR}$ Xeon$^{\circledR}$ processor 3.7 GHz and 16 GB RAM. The DI-EPA framework was programmed in MATLAB R2019a using the YALMIP toolbox~\cite{Lofberg2004} and solved using the GUROBI solver~\cite{gurobi}.

\begin{table}[tbp]
    \centering
    \caption{Case Study Parameters}
    \vspace{-1mm}
    \begin{tabular}{l|c c}
    \hline
    Parameter & Unit & Value\\
    \hline
    Battery Capacity & $\rm kWh$ & 500 \\
    Charge/Discharge efficiency & p.u. & 1\\
    No. of Samples $|\cal S|$ & - & 250 \\
    Min. Expected Profit $\varphi^{\star}$ & $\$$ & 1 \\
    Degradation Coefficient $a_1$ & - & 5.24e-4 \\
    Degradation Coefficient $a_2$ & - & 2.03 \\
    \hline
    \end{tabular}
    \label{table:data}
\end{table}

\begin{figure}[tbp]
\centering
\begin{subfigure}{0.95\columnwidth}
    \includegraphics[width=\textwidth]{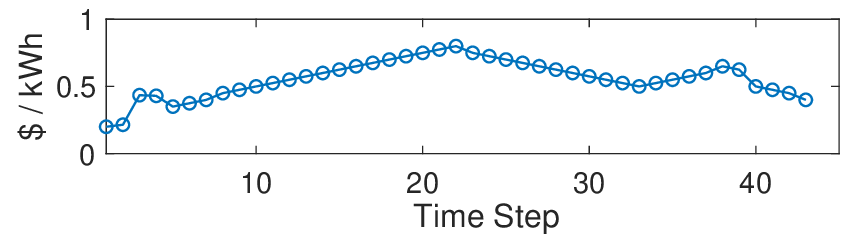}\vspace{-1mm}
    \caption{Mean energy prices}
    \label{fig:rate_en}
\end{subfigure}
\vfill
\begin{subfigure}{0.95\columnwidth}
    \includegraphics[width=\textwidth]{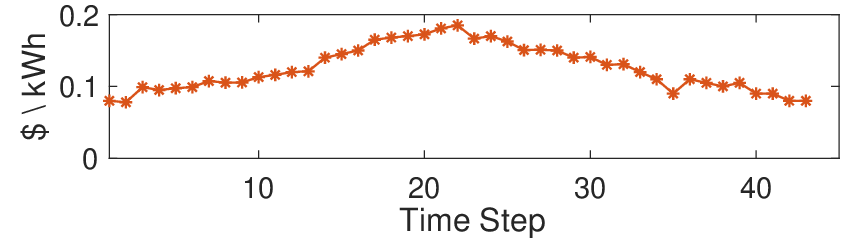}\vspace{-1mm}
    \caption{Mean reserve prices}
    \label{fig:rate_res}
\end{subfigure}
\caption{Energy and reserve prices}
\label{fig:rate}
\vspace{-5mm}
\end{figure}






\vspace{-2mm}
\subsection{\textcolor{black}{Comparison with the Risk-Neutral Approach}}
To analyze the effectiveness of the proposed DI-EPA framework in handling uncertain market prices, the given case study was simulated using two approaches. In the first approach, the facilitator uses \textcolor{black}{a risk-neutral approach to optimize the ESS schedules}. In this approach, the optimization is done over each sample of price independently and the power dispatch is averaged out for each interval at the end. The resulting power dispatch is then used to estimate the expected profit over the same set of price samples. In the second approach, the facilitator uses the proposed method with a confidence level of $95 \%$. The results are compared in Table \ref{tab:MC}. It can be observed that the proposed method provides a slightly improved profit and requires significantly less computation time (44.6 seconds). Further, it was observed that the minimum expected profit  $\zeta^{\star}_{\alpha}$ determined by the proposed DI-EPA framework (computed based on ${\rm VaR}_{0.95}$) was equal to or higher than $\$ 184.8$ for $95\%$ of the samples whereas the corresponding profit value obtained from the first approach was estimated to be $\$175.32$. Thus, the proposed approach leads to 5.13 \% lower risk compared to the risk-neutral approach.

As the facilitator consists of three ESSs, there exists $8$ $(=2^3)$ coalitions for their cooperative participation in markets, i.e., \{$\varnothing$\}, \{ESS~1\}, \{ESS~2\}, \{ESS~3\}, \{ESS~1, ESS~2\}, \{ESS~2, ESS~3\}, \{ESS~3, ESS~1\}, and \{ESS~1, ESS~2, ESS~3\}. Here, \{$\varnothing$\} means no ESS in the facilitator, and hence, it results in zero profit by default. Considering the ${\rm VaR}_{0.95}$, i.e., the minimum expected profit with $95\%$ confidence, it was observed that adding each ESS resulted in an improvement in profit leading to the highest profit obtained at the grand coalition. Fig.~\ref{fig:shapley_val} depicts the contribution of each ESS towards VaR and CVaR at a $95\%$ confidence level. Accordingly, the profit proportions are decided such that each ESS is fairly compensated from the actual profit resulting in operation.
\begin{table}[tbp]
 \caption{Comparison of profit distribution with the proposed method and the Monte Carlo based periodic optimization method}
    \centering
    \begin{tabular}{l|c| c| c}
        & Exp. Profit    & Profit at $95 \%$    & Computation \\
        & ($\$ $)  & Confidence ($\$ $)& Time ($ \rm sec. $)  \\
        \hline
    Risk-neutral approach  & 263.91  & 175.32  &  337.6 \\
    Proposed ($\alpha = 0.95)$ & 266.04 & 184.80  & 44.6  \\
     \hline
    \end{tabular}
    \label{tab:MC}
\end{table}
\begin{figure}[tbp]
\centering
\includegraphics[width = 0.9\columnwidth]{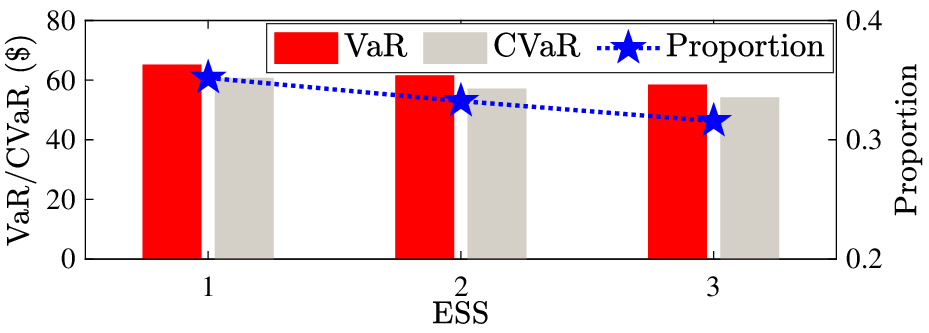}\vspace{-2mm}
\caption{VaR, CVaR, and profit proportion of each ESS considering Shapley Value.} 
\vspace{-5mm}
\label{fig:shapley_val}
\end{figure}

\color{black}
\vspace{-4mm}
\subsection{Sensitivity towards Confidence Level}
For various confidence levels $\alpha$, Table~\ref{tab:risk_level} reports the expected profit, VaR $\zeta^{\star}_{\alpha}$, and CVaR $\tilde{\cal O}^{\star}_{\alpha}$ values in addition to the amount of energy dispatch. Therein, it was observed that the expected profit, VaR, and CVaR values are reduced with the increased confidence level $\alpha$. As mentioned in Section~\ref{sec:cvar}, a higher $\alpha$ leads to a lower expected profit which is less risky and vice versa. It is also evident from the energy dispatch values that the battery is subjected to heavy usage (multiple charge/discharge cycles) when high risk is permitted or when $\alpha$ is low. From the VaR and CVaR figures in Table~\ref{tab:risk_level}, it is observed that there is a $0.05$ probability (i.e., at $\alpha=0.95$) that the profit drops below $\$ 184.8$, and in such an undesirable situation, the profit is on average $\$171.66$. Similar information can be realized from the respective figures at the other confidence levels as well.

\begin{table}[tbp]
\color{black}
 \caption{\textcolor{black}{Profit and Dispatch Variation Against $\alpha$}}
    \centering
    \begin{tabular}{l|c c c c}
    \hline
     Confidence Level $\alpha$    &  0.75  & 0.85  &  0.95    & 0.99 \\ \hline
     Expected Profit ($\$ $)  &  275.49  &  270.58 & 266.04 & 266 \\
     VaR $\zeta$ ($\$ $) & 236.46 & 211.39  & 184.80  & 163.19 \\
     CVaR ($\$ $) &  203.52  & 188.58  & 171.66 &  158.15 \\ \hline
     Energy Dispatched$^\#$ ($\rm kWh$) &  4360.6   &  3938.8   & 3687.8   & 3602 \\
     \hline
     \multicolumn{5}{l}{$^\#$ Energy Dispatched is defined as $\sum^B_{j=1}\boldsymbol{1}^\top|D_{j}|\boldsymbol{1}$}
    \end{tabular}
    \label{tab:risk_level}
\end{table}

The cumulative distribution function of the profit at different confidence levels is depicted Fig.~\ref{fig:risk_level}. It is observed that the profit distribution and the expected profit varies with the confidence level due to the change in the optimal dispatch profile computed by the DI-EPA algorithm.

\begin{figure}[tbp]
    \centering
    \includegraphics[width= 0.95\columnwidth]{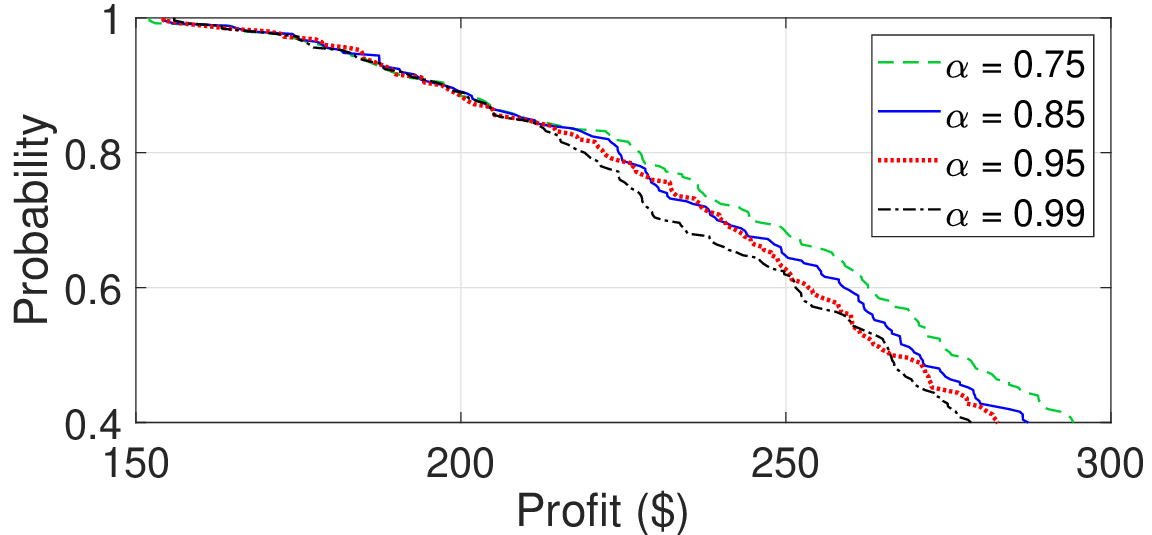}
    \caption{\textcolor{black}{Cumulative distribution of profit at various confidence levels}}
    \vspace{-2mm}
    \label{fig:risk_level}
\end{figure}
 
\subsection{Sensitivity towards Number of Samples}
To analyze the effect of the number of training samples taken during optimization, the simulations were performed with $250$ and $1000$ samples while keeping the confidence level the same ($95 \%$). The results are shown in Table~\ref{tab:accuracy}. It can be observed that the expected profit values in the two cases are comparable. The respective discrepancies between VaR and CVaR values are less than $3\%$. 
Although the results with 1000 samples depict a higher accuracy, the computation time is almost $2.76$ times higher than that with 200 samples.

\begin{table}[tbp]
\color{black}
 \caption{\textcolor{black}{Accuracy Analysis at $95\%$ confidence level}}
    \centering
    \begin{tabular}{c| c| c| c| c}
    \hline
     \# Samples   & Exp. Profit ($\$ $)   &  VaR ($\$ $)   & CVaR ($\$ $)  & Comp. time (s)\\
        \hline
    250 & 266.04 & 184.80 & 171.66 & 44.6\\
    1000 & 268.77 & 179.15 & 166.95 & 123.3\\
     \hline
    \end{tabular}
    \label{tab:accuracy}
\end{table}

\color{black}
\subsection{Effect of Considering COD over ESS Dispatch}
To analyze the effects of including COD in the scheduling strategy, the DI-EPA framework was executed over a total of 10 hours consisting of $40$ market intervals. This is established by incorporating the DI-EPA framework into a model predictive control-based optimization framework with a prediction horizon of four $15$-minute intervals. The battery dispatch scheduling results with and without including the COD of batteries in the DI-EPA framework are shown in Table~\ref{tab:MPC}. The table shows the expected revenue earned through energy arbitrage, monetary loss due to degradation, and net profit for each battery. It can be observed that all the batteries undergo a lower amount of capacity degradation when degradation cost is included in decision-making via the DI-EPA framework. Since the replacement cost for ESS $3$ was the highest among all the three ESSs, a lesser amount of energy is dispatched from ESS~$3$ leading to its lower expected revenue and a lower degradation cost. \textcolor{black}{Further, the expected net profit is observed to be approximately $5.64 \%$ higher for the case when COD is included in comparison to the case that omits COD within the decision-making process.}

\begin{table}[tbp]
\centering
  \caption{Net Profit over 40 intervals with \& without including degradation costs in the proposed EPA Formulation}\vspace{-1mm}
    \begin{subtable}[t]{0.45\textwidth}
        \centering
         \caption{Scheduling with degradation costs included}\vspace{-1mm}
        \begin{tabular}{l | c | c | c}
        \hline
         & Exp. Revenue & Deg. Cost & Exp. Net Profit \\
        \hline  
        ESS 1 & \textcolor{black}{481.87} & 47.79 & \textcolor{black}{434.08} \\
        ESS 2 & \textcolor{black}{263.02}  & 24.74 & \textcolor{black}{238.28} \\
        ESS 3 & \textcolor{black}{241.50}  & 23.13 & \textcolor{black}{218.37} \\
        \hline
        Total & \textcolor{black}{986.39}  & 95.66 & \textcolor{black}{890.73} \\
        \hline
       \end{tabular}
       \label{tab:mpc1}
    \end{subtable}
    \vfill
    \vspace{1em}
    \begin{subtable}[t]{0.45\textwidth}
        \centering
         \caption{Scheduling without degradation costs included}\vspace{-1mm}
        \begin{tabular}{l | c | c | c}
        \hline
         & Exp. Revenue & Deg. Cost & Exp. Net Profit \\
        \hline  
        ESS 1 & \textcolor{black}{311.21}  & 57.22 & \textcolor{black}{253.99} \\
        ESS 2 & \textcolor{black}{372.92}  & 77.84 & \textcolor{black}{295.08} \\
        ESS 3 & \textcolor{black}{372.49}  & 78.37 & \textcolor{black}{294.12} \\
        \hline
        Total & \textcolor{black}{1056.62}  & 213.43 & \textcolor{black}{843.19} \\
        \hline
       \end{tabular}
       \label{tab:mpc2}
     \end{subtable}
   \vskip -2em
     \label{tab:MPC}
\end{table}

\section{Conclusion}\label{sec:conclution}
The paper presents a degradation-infused fair sharing economy-based framework for the participation of energy storage in multi-market environments. Considering the uncertainty in market prices of energy and reserve, the facilitator solves the proposed energy portfolio allocation problem formulated as a multi-objective optimization problem that jointly maximizes the expected profit and minimizes the associated risk for a given confidence level. The profit is expressed as the difference between revenue earned from the market and the capacity loss due to degradation. A closed-form degradation function was obtained using the Gaussian process. The sensitivity of results towards the number of samples and confidence level was analyzed. The results illustrate that increasing the confidence level in the proposed optimization problem brings down the value-at-risk and the conditional value-at-risk effectively. Then, using the Shapley value concept, the benefits of each coalition are comprehensively studied, and fair profit allocations to individual ESSs are determined based on their contribution. \textcolor{black}{Later, the effect of considering the degradation cost in the proposed framework was analyzed and it was found that the expected net profit value was 5.64 \% higher as the proposed method optimizes the battery schedules based on their replacement costs. The proposed method was also found to be having 5.13 \% lower value at risk at 95 \% confidence level compared to a risk-neutral approach.}

\bibliographystyle{IEEEtran}

\end{document}